%% file: main.tex
\documentclass[sigconf]{acmart}

\usepackage{multirow}
\usepackage{booktabs}
\usepackage{graphicx}
\usepackage{subcaption}

\usepackage{caption}
\AtBeginDocument{%
  }

\copyrightyear{2026}
\acmYear{2026}
\setcopyright{cc}
\setcctype{by}
\acmConference[SIGIR '26]{Proceedings of the 49th International ACM SIGIR Conference on Research and Development in Information Retrieval}{July 20--24, 2026}{Melbourne, VIC, Australia}
\acmBooktitle{Proceedings of the 49th International ACM SIGIR Conference on Research and Development in Information Retrieval (SIGIR '26), July 20--24, 2026, Melbourne, VIC, Australia}
\acmDOI{10.1145/3805712.3809929}
\acmISBN{979-8-4007-2599-9/2026/07}

\begin{document}

\title{Every Preference Has Its Strength: Injecting Ordinal Semantics into LLM-Based Recommenders}

\author{Jiwon Jeong}
\email{zzioni@kaist.ac.kr}
\affiliation{%
  \institution{Korea Advanced Institute of Science and Technology (KAIST)}
  \city{Daejeon}
  \country{Republic of Korea}
}

\author{Donghee Han}
\email{handonghee@kaist.ac.kr}
\affiliation{%
  \institution{Korea Advanced Institute of Science and Technology (KAIST)}
  \city{Daejeon}
  \country{Republic of Korea}
}

\author{Sungrae Hong}
\email{sr5043@kaist.ac.kr}
\affiliation{%
  \institution{Korea Advanced Institute of Science and Technology (KAIST)}
  \city{Daejeon}
  \country{Republic of Korea}
}

\author{Woosung Kang}
\email{wskang@kaist.ac.kr}
\affiliation{%
  \institution{Korea Advanced Institute of Science and Technology (KAIST)}
  \city{Daejeon}
  \country{Republic of Korea}
}

\author{Mun Yong Yi}
\authornote{Corresponding author.}
\email{munyi@kaist.ac.kr}
\affiliation{%
  \institution{Korea Advanced Institute of Science and Technology (KAIST)}
  \city{Daejeon}
  \country{Republic of Korea}
}


\renewcommand{\shortauthors}{Jiwon Jeong, Donghee Han, Sungrae Hong, Woosung Kang, \& Mun Yong Yi}

\begin{abstract}
Recent work has shown that large language models (LLMs) can enhance recommender systems by integrating collaborative filtering (CF) signals through hybrid prompting. However, most existing CF–LLM frameworks collapse explicit ratings into implicit or positive-only feedback, discarding the ordinal structure that conveys fine-grained preference strength. As a result, these models struggle to exploit graded semantics and nuanced preference distinctions.
We propose \textbf{O}rdinal \textbf{S}emantic \textbf{A}nchoring (\textbf{OSA}), a hybrid CF–LLM framework that explicitly incorporates preference strength by modeling \emph{interaction-level} user feedback. OSA represents ordinal preference levels as numeric textual tokens and uses their token embeddings as semantic anchors to align user–item interaction representations in the LLM latent space. Through strength-aware alignment across ordinal levels, OSA preserves preference semantics when integrating collaborative signals with LLMs.
Experiments on multiple real-world datasets demonstrate that OSA consistently outperforms existing baselines, particularly in pairwise preference evaluation, highlighting its effectiveness in modeling fine-grained user preferences over prior CF--LLM methods.

\end{abstract}


\maketitle

\input{texs/1.Introduction.tex}
\input{texs/2.Method.tex}

\input{texs/3.Experiment.tex}
\input{texs/4.Conclusion.tex}

\begin{acks}
This work was supported by the National Research Foundation of Korea(NRF) grant funded by the Korean government(MSIT) (No.RS-2022-NR068758).
\end{acks}

\bibliographystyle{ACM-Reference-Format}
\bibliography{main}

\end{document}

%% file: texs/1.Introduction.tex
\section{Introduction}

\begin{figure}[t]
  \centering
  \includegraphics[width=\columnwidth]{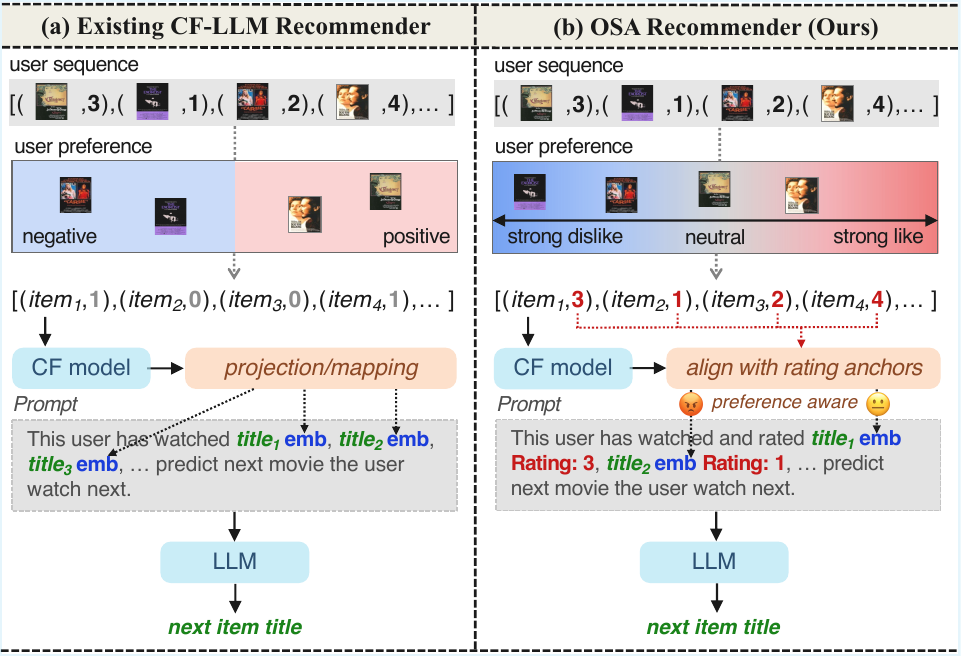}
  \vspace{-5mm}
  \caption{Comparison between existing CF–LLM recommenders and the proposed OSA framework.}
  \label{fig:osa}
\end{figure}

Recent work has shown that large language models (LLMs) can significantly improve recommender systems by leveraging rich textual information and strong generalization capabilities~\cite{llm4rec,llmrec1,llmrec2,llmrec3,llmrec4}.
A prominent line of this research adopts hybrid prompting~\cite{llara, collm, allmrec}, which integrates collaborative filtering (CF) signals into the LLM’s input or latent space.
This integration is typically achieved by projecting or aligning user or item representations from CF models within the LLM, allowing the LLM to jointly leverage textual semantics and collaborative information. 
Most existing approaches frame this integration primarily as a representation mapping or projection problem~\cite{cora, ilora}, for example by injecting embeddings from CF models as additional tokens~\cite{collm,llara} or applying distillation~\cite{dllm2rec} and binarization~\cite{binllm} techniques. 
These works have shown that injecting collaborative information into LLM-driven recommendation can yield substantial improvements over purely textual~\cite{llmrank, tallrec} or purely collaborative baselines~\cite{grurec,caser,sasrec}.

Despite these advances, current hybrid LLM–CF integration frameworks often overlook fine-grained user preference signals, such as explicit ratings, which are essential for capturing nuanced individual tastes.
Rather than serving as mere numeric labels, ratings encode both whether a user liked or disliked an item and the strength of that preference. 
For instance, a 5-star rating reflects a substantially stronger preference than a 4-star rating, while a 1-star rating indicates stronger dislike than a 2-star rating. As such, ratings provide an ordinal semantic signal that captures fine-grained variations in preference intensity.
However, in most hybrid LLM–CF approaches, ratings are commonly binarized using a rating threshold~\cite{llara,allmrec} or reduced to positive-only interactions~\cite{collm, binllm}. 
As a result, these models tend to capture coarse preference signals that suffice for identifying generally liked items, but are less effective at modeling the fine-grained preference strength required to resolve subtle distinctions among closely related candidates.

While rating simplification has been a reasonable design choice in existing recommender systems~\cite{binarize}, it becomes a suboptimal one when LLMs are employed as recommendation predictors. 
LLMs are particularly well suited to modeling graded semantics and nuanced comparisons, and can therefore benefit from supervision signals that preserve fine-grained preference information~\cite{llmpref,setwise,multipref}. 
When ratings are reduced to implicit or positive-only feedback, this opportunity is largely missed, and the LLM is used as a generic predictor rather than as a semantic model.


To overcome these challenges (Figure~\ref{fig:osa}), we propose \textbf{O}rdinal \textbf{S}emantic \textbf{A}nchoring \textbf{(OSA)}, a hybrid CF–LLM framework that incorporates fine-grained user preference strength into LLM-based recommender using explicit ratings. 
OSA represents rating levels as numeric textual tokens and uses their token embeddings as semantic anchors to align \emph{interaction-level} user–item representations from a CF model in the LLM latent space. 
During training, OSA applies strength-aware alignment in proportion to preference intensity, pulling interactions more strongly toward anchors for extreme ratings while maintaining separation across ordinal levels.
This anchoring introduces an inductive bias that preserves the ordinal structure of preference strength when integrating collaborative representations with the LLM, enabling OSA to capture not only whether an item is preferred, but also how strongly it is preferred.

Our contributions are as follows: 

\noindent(1) We identify ordinal semantics in explicit ratings as an overlooked  objective in hybrid LLM–CF recommendation, highlighting the importance of preserving fine-grained user preferences. 

\noindent(2) We propose Ordinal Semantic Anchoring, which enables strength-aware alignment by grounding interaction representations to rating-level semantic anchors within the LLM latent space. 

\noindent(3) We empirically and qualitatively show that this anchored alignment improves both fine-grained preference modeling and overall recommendation performance.

%% file: texs/2.Method.tex
\section{Method}

\begin{figure}[t]
  \centering
  \includegraphics[width=\linewidth]{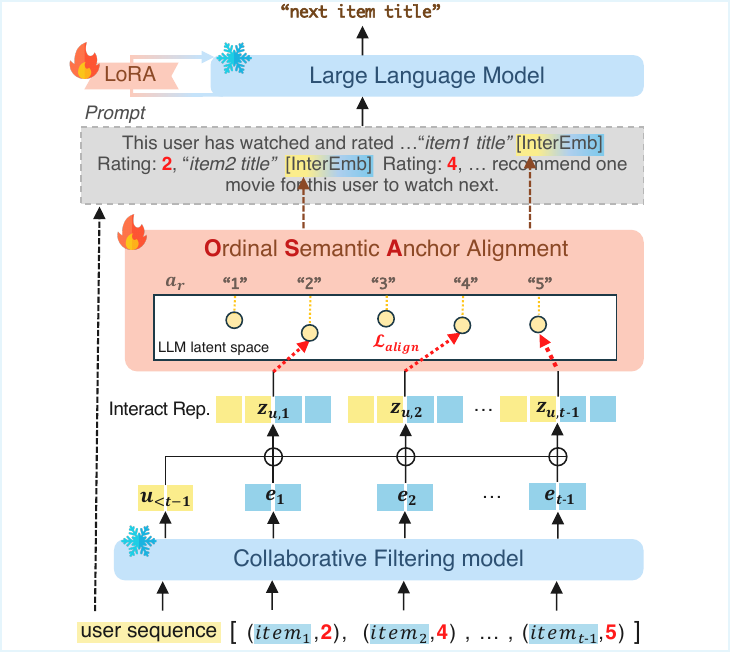}
  \vspace{-5mm}
  \caption{Overview of Ordinal Semantic Anchoring (OSA).}
  \label{fig:osa_overview}
\end{figure}

We propose \textbf{O}rdinal \textbf{S}emantic \textbf{A}nchoring \textbf{(OSA)}, a CF--LLM hybrid framework for sequential recommendation that incorporates fine-grained preference strength by aligning interaction representations with ordinal semantic anchors derived from explicit ratings. As illustrated in Figure~\ref{fig:osa_overview}, OSA operates under a next-item
prediction setting, where the model predicts the most likely next item based on
a user's interaction history and a set of candidate items.

\subsection{Interaction Representation}

OSA builds on a standard CF model (e.g., SASRec) to encode user--item
interactions.
For each user $u \in \mathcal{U}$, let $\mathcal{H}_u^{<t} = (i_1, \dots, i_{t-1})$
denote the interaction history up to time $t-1$, where $i_k \in \mathcal{I}$.
The CF encoder maps $\mathcal{H}_u^{<t}$ to a latent user representation
\begin{equation}
\mathbf{u}_{<t} = f_{\text{CF}}\!\left(\mathcal{H}_u^{<t}\right)
\in \mathbb{R}^{d_{\text{rec}}},
\end{equation}
which captures the user’s preference context prior to time $t$.

The CF model also maintains a shared item embedding table learned across all users,
where each item $i$ is associated with an embedding
\begin{equation}
\mathbf{e}_i \in \mathbb{R}^{d_{\text{rec}}},
\end{equation}
which encodes item-level collaborative signals shared across users.

We construct an interaction representation $\mathbf{z}_{u,i}$ by concatenating the
user and item representations,
\begin{equation}
\mathbf{z}_{u,i} = [\mathbf{u}_{<t-1} \, ; \, \mathbf{e}_i] \in \mathbb{R}^{2d_{\text{rec}}},
\end{equation}
which corresponds to user $u$’s interaction with item $i$ at time $t_{u,i}$. Each $\mathbf{z}_{u,i}$
is associated with an explicit rating signal $r_{u,i} \in \{1,\dots,5\}$, reflecting
the observed preference strength. By modeling representations at the level of interactions,
rather than at the user or item level, OSA establishes a direct correspondence between interaction-level representations and the individual preference expressed in each interaction.

\subsection{Projector and Hybrid Prompting}

To interface CF representations with the LLM, OSA applies a learnable projector
\begin{equation}
f_\text{proj}: \mathbb{R}^{2d_{\text{rec}}} \rightarrow \mathbb{R}^{d_{\text{llm}}},
\end{equation}
implemented as a two-layer MLP with GELU. For each interaction representation
$\mathbf{z}_{u,i}$, we obtain a projected vector $\mathbf{v}_{u,i}=f_\text{proj}(\mathbf{z}_{u,i})$,
which is used as an input to the LLM.

\begin{figure}[t]
  \centering
  \includegraphics[width=\linewidth]{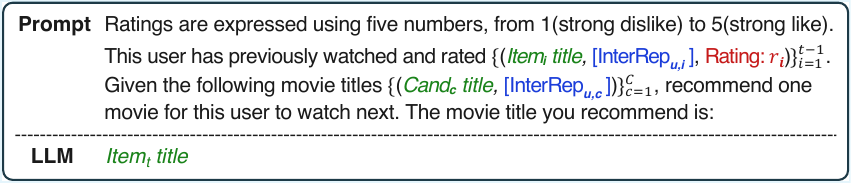}
  \vspace{-6mm}
  \caption{Prompt template for OSA.}
  \label{fig:prompt}
\end{figure}

We adopt a pretrained LLM and fine-tune it with LoRA~\cite{lora} to adapt the model to the
recommendation task. Following the template in Figure~\ref{fig:prompt}, the model receives a hybrid prompt with the user's rated interaction history and a candidate set $\mathcal{C}=\{c_1,\dots,c_C\}\subseteq\mathcal{I}$, augmented with $\mathbf{v}_{u,i}$.
Given this input, the LLM outputs the next-item title from the candidate set.

\subsection{Ordinal Semantic Anchoring}

In prior CF--LLM hybrid methods, preference information from explicit ratings
is absorbed into implicit CF signals, leaving the LLM without access to fine-grained preference signals during recommendation.
By contrast, OSA preserves this information by modeling ratings as \emph{ordinal semantic
anchors} in the LLM latent space and explicitly aligning interaction
representations to these anchors.

Specifically, each rating level $r \in \{1,\dots,5\}$ can be represented as a numeric
textual token (e.g., ``1'', ``2'', \dots, ``5''). We extract the corresponding token
embeddings from the pretrained LLM and treat them as fixed anchor vectors
\begin{equation}
\mathbf{a}_r \in \mathbb{R}^{d_{\text{llm}}} \qquad \forall\, r \in \{1,\dots,5\}.
\end{equation}
These anchors leverage the LLM’s prior semantic understanding of ordered numeric scales and provide a shared, interpretable reference for collaborative representations.

\subsection{Strength-Aware Alignment}

Given a projected interaction representation $\mathbf{v}_{u,i}$ and its rating
$r_{u,i}$, OSA aligns $\mathbf{v}_{u,i}$ with the corresponding ordinal anchor
$\mathbf{a}_{r_{u,i}}$, whose semantics are grounded in the LLM’s inherent
understanding.
Through this \emph{semantic-grounded alignment}, interaction representations are
structured with respect to the ordinal meaning conveyed by ratings, allowing
collaborative signals to be interpreted within the LLM’s semantic space. This grounding makes the CF representations more semantically interpretable to the LLM, guiding it to leverage collaborative signals in a preference-aware manner.

Aligning interaction-level representations to their observed preference strength helps capture more individual tastes than relying solely on global CF representations.
As a result, preference distinctions are preserved in a fine-grained manner,
with interactions positioned to respect the ordinal structure induced by
the anchors in the LLM latent space.

We formalize this objective as an alignment loss:
\begin{equation}
\mathcal{L}_{\text{align}} =
\mathbb{E}_{(u,i)} \Big[ \big(1 - \cos(\mathbf{v}_{u,i}, \mathbf{a}_{r_{u,i}})\big)
\cdot w(r_{u,i}) \Big],
\end{equation}
where $\cos(\cdot,\cdot)$ denotes cosine similarity. To reflect the intuition that extreme ratings express stronger and more definitive human preference signals, we introduce
a strength-aware weight
\begin{equation}
w(r) = 1 + \gamma \cdot |r - 3|,
\end{equation}
with $\gamma > 0$. This design encourages interactions associated with strong likes
or dislikes to align more tightly with their corresponding anchors, while allowing
neutral interactions to remain less constrained. Through this strength-aware alignment, OSA organizes interaction representations
in the LLM latent space to follow an ordinal semantic structure of preference
intensity, rather than collapsing them into coarse similarity groups.

Following prior CF--LLM hybrid recommenders \cite{llara, allmrec}, we train the model to generate the
ground-truth next-item title from a structured prompt augmented with
collaborative signals.
Formally, let $\mathbf{p}_u$ be the input prompt of user $u$ and $\mathbf{y}_{u,i_{t}}$ be the
target item title. The autogressive objective is
\begin{equation}
\max_{\theta}\ \sum_{u \in \mathcal{U}} \sum_{k=1}^{|\mathbf{y}_{u, i_{t}}|}
\log P_{\theta,\Theta}\!\left(
y_{u, i_{t}}^{(k)} \mid \mathbf{p}_u, \mathbf{y}_{u, i_{t}}^{(<k)}
\right),
\end{equation}
where $\theta$ denotes learnable parameters (projector and LoRA) and $\Theta$
is the frozen LLM.

%% file: texs/3.Experiment.tex
\begin{table}[!t]
\centering
\small
\renewcommand{\arraystretch}{1.0} 
\setlength{\tabcolsep}{6pt}
\caption{Dataset statistics.}
\begin{tabular}{lccc}
\hline
\textbf{} & \textbf{ML-1M} & \textbf{Scientific} & \textbf{Games} \\
\hline
\textbf{\# Interactions} & 999,917 & 179,712 & 568,508 \\
\textbf{\# Users}        & 6,040   & 28,037  & 64,073  \\
\textbf{\# Items}        & 3,503   & 28,461  & 33,614  \\
\hline
\end{tabular}
\label{tab:data_stats}
\end{table}

\section{Experiments}

\subsection{Experimental Settings}

\textbf{Datasets.}
We conduct experiments on three real-world datasets, MovieLens-1M (ML-1M)~\cite{movielens} and two Amazon datasets~\cite{amazon}, Industrial and Scientific (Scientific) and Video Games (Games). All datasets provide explicit 1--5 star ratings. As summarized in Table~\ref{tab:data_stats}, the datasets vary substantially
in size, enabling evaluation across different data scales.
\\
\textbf{Evaluation Protocol.}
We evaluate OSA under a next-item prediction, where the model observes
up to 20 historical interactions and predicts the next item from one
ground-truth item and 19 randomly sampled negatives.
We adopt a leave-one-out split, using the most recent interaction for
testing, the second most recent for validation, and the rest for training. We report Hit@1 as the primary metric, since the task is formulated
as a single-choice selection from the candidate set. To further assess fine-grained preference modeling, we conduct a pairwise evaluation by sampling $\min(1000, 0.05|\mathcal{U}|)$ unseen users and constructing one-level rating pairs, categorized as strong (1 vs.\ 2, 4 vs.\ 5) and subtle (2 vs.\ 3, 3 vs.\ 4).
\\
\textbf{Baselines.}
We compare OSA with three categories of baselines.
Collaborative sequential recommendation models include
SASRec~\cite{sasrec}, GRU4Rec~\cite{grurec}, and Caser~\cite{caser}.
We also evaluate LLM-only recommenders that rely solely on textual inputs,
including TALLRec~\cite{tallrec} and LLMRank~\cite{llmrank}.
Finally, we compare with existing hybrid CF--LLM recommenders,
LLaRA~\cite{llara} and A-LLMRec~\cite{allmrec},
which integrate collaborative filtering signals into LLMs.
\\
\textbf{Settings.}
All LLM-based models use LLaMA 3.2 (3B-Instruct) as a common backbone, and SASRec is used as the shared CF encoder for all hybrid baselines. We set $d_{\text{rec}}=64$, $\lambda_{\text{align}}=0.5$, and $\gamma=0.5$.
Training settings and implementation details are provided in our code repository at \url{https://github.com/zzzzzioni/OSA}.

\subsection{Overall Performance}
Table~\ref{tab:main_results} reports Hit@1 performance on three datasets.
In most cases, hybrid CF--LLM models outperform CF-only and LLM-only baselines,
indicating the benefit of combining collaborative signals with LLMs for
recommendation.
Beyond the general benefit of hybrid approaches, OSA further improves performance
by explicitly modeling fine-grained preference strength at the level of
individual user--item interactions. By aligning interaction representations with rating-derived ordinal structure
in a manner consistent with the semantic space of the LLM, OSA allows the model
to interpret CF representations as semantically meaningful signals during
recommendation. This design leads to consistent improvements in recommendation performance across all
datasets. 

\begin{table}[t]
\centering
\small
\renewcommand{\arraystretch}{1.0}
\setlength{\tabcolsep}{6pt}
\caption{Hit@1 performance under next item prediction.
The best result for each dataset is highlighted in bold.}
\begin{tabular}{l lccc}
\toprule
\textbf{Model} & & \textbf{ML-1M} & \textbf{Scientific} & \textbf{Games} \\
\midrule
\multirow{3}{*}{\centering CF-only}
 & GRU4Rec  & 0.4677 & 0.2207 & 0.4090 \\
 & Caser    & 0.5168 & 0.2601 & 0.4593 \\
 & SASRec   & 0.5269 & 0.2664 & 0.5216 \\
\midrule
\multirow{2}{*}{\centering LLM-only}
 & LLMRank  & 0.0464 & 0.0487 & 0.0528 \\
 & TALLRec  & 0.4032 & 0.0294 & 0.4568 \\
\midrule
\multirow{3}{*}{\centering Hybrid}
 & LLaRA    & 0.5816 & 0.3404 & 0.5835 \\
 & A-LLMRec & 0.4920 & 0.3606 & 0.5082 \\
 & \textbf{OSA(Ours)} 
 & \textbf{0.6014} & \textbf{0.4338} & \textbf{0.6326} \\
\bottomrule
\end{tabular}
\label{tab:main_results}
\end{table}

\begin{figure}[t]
  \centering
  \includegraphics[width=0.97\linewidth]{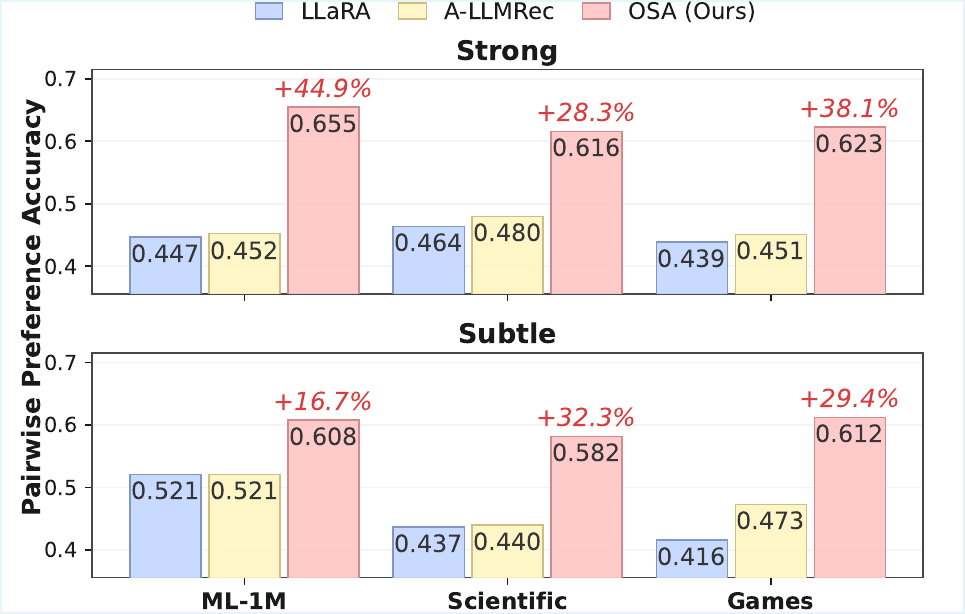}
  \vspace{-2mm}
  \caption{Pairwise preference accuracy (Strong and Subtle),
with relative gains (+\%) over the best baseline.}
  \label{fig:pairwise_bar}
\end{figure}

\subsection{Pairwise Preference Evaluation}
To assess fine-grained preference distinctions beyond prior hybrid methods, we additionally report pairwise
preference accuracy on unseen users. Following the format in Figure 3, we modify the instruction to: \texttt{"Given the following 2 product names: [Cands], choose the one this user would like more and give a higher rating."}
As shown in Figure~\ref{fig:pairwise_bar}, OSA substantially outperforms LLaRA and
A-LLMRec across all datasets and both strong and subtle settings, with margins
relative improvements reaching up to 44.9\%. Binarizing explicit ratings discards the fine-grained information needed to discern subtle preference distinctions between items. Consequently, baselines often converge to 0.5, performing no better than a random guess.
This result highlights that preserving ordinal preference strength at the
interaction level is critical for distinguishing subtle preference differences
that are essential for accurately understanding user intent in recommendation.
Such fine-grained preference signals are difficult to capture in existing hybrid
methods, where ratings are often discarded or reduced to implicit feedback.

\subsection{Ablation Study}
We conduct the following ablation studies to assess the contribution of individual components in OSA using Hit@1.
(i) \textbf{w/o $\gamma$}: removing strength-aware weighting;
(ii) \textbf{w/o $\mathcal{L}_{\text{align}}$}: removing the alignment objective;
(iii) \textbf{Item-only Rep.}: replacing $\mathbf{z}_{u,i}$ with $\mathbf{e}_i$;
(iv) \textbf{w/o ordinal structure}: permuting rating anchors to break their ordinal ordering; and
(v) \textbf{w/o semantics}: replacing rating anchors with randomly initialized vectors.
Table~\ref{tab:ablation} presents the ablation results.
Removing interaction-level representations or disrupting ordinal anchors leads to clear performance drops, confirming the importance of fine-grained modeling and ordinal semantic structure.
Performance also degrades without $\gamma$, indicating the benefit of emphasizing strong preferences.
Finally, removing the alignment objective reduces performance toward hybrid baselines, underscoring the necessity of anchored alignment.

\begin{table}[t]
\centering
\small
\renewcommand{\arraystretch}{1.0}
\caption{Ablation study of OSA components.}
\begin{tabular}{lccc}
\toprule
\textbf{Ablation} & \textbf{ML-1M} & \textbf{Scientific} & \textbf{Games} \\
\midrule
w/o $\mathcal{L}_{\text{align}}$        & 0.5750 & 0.3941 & 0.5654 \\
w/o $\gamma$  & 0.5969 & 0.4155 & 0.5769 \\
Item-only Rep.        & 0.5630 & 0.3310 & 0.6082 \\
w/o ordinal structure & 0.5583 & 0.3887 & 0.5578 \\
w/o semantics & 0.5515 & 0.3959 & 0.5423 \\
\midrule
\textbf{OSA}       & \textbf{0.6014} & \textbf{0.4338} & \textbf{0.6326} \\
\bottomrule
\end{tabular}
\label{tab:ablation}
\end{table}

\begin{figure}[t]
  \centering
  \includegraphics[width=0.97\linewidth]{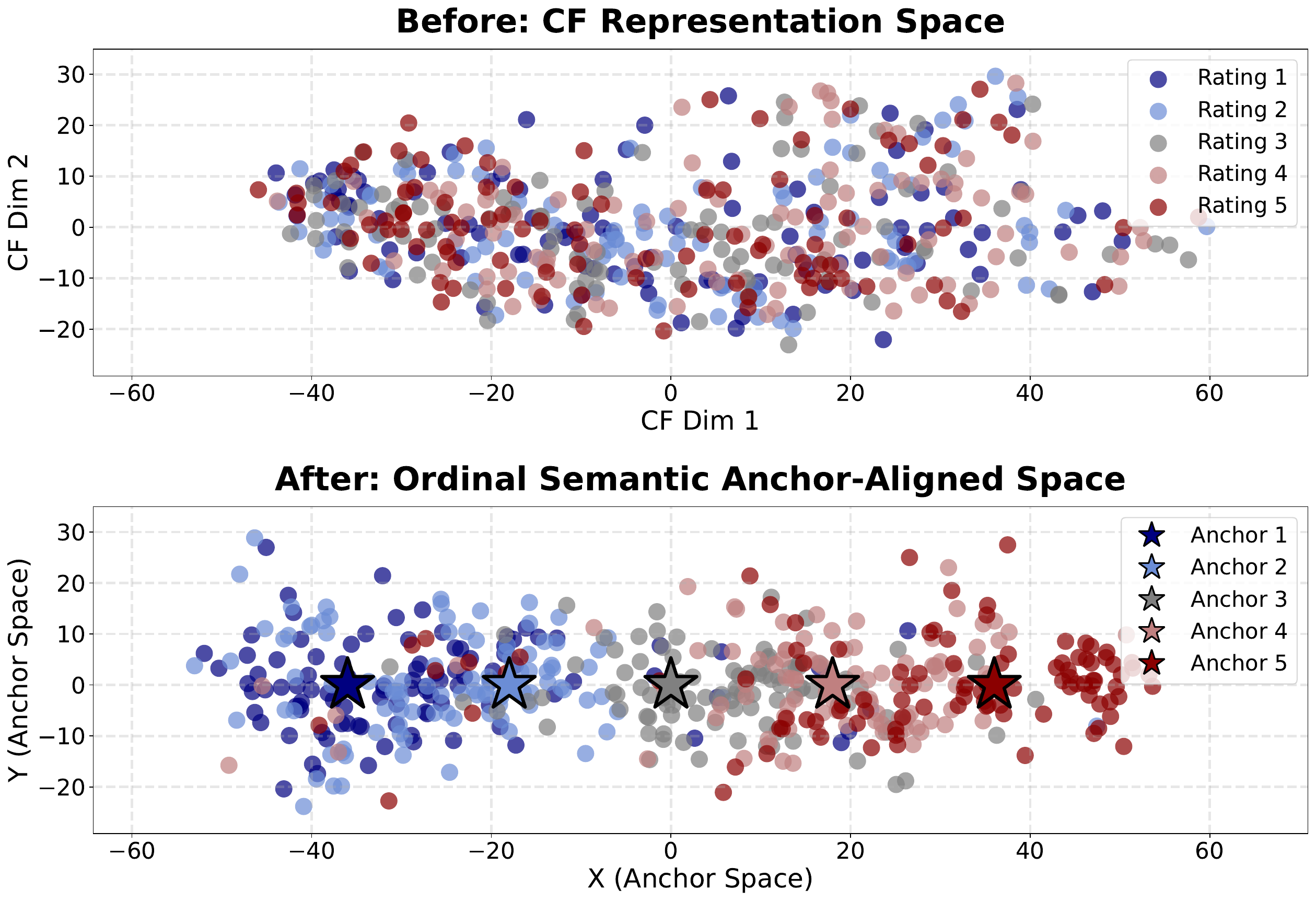}
  \vspace{-2mm}
    \caption{Visualization of interaction representations before and after applying ordinal semantic anchoring (ML-1M).}
  \label{fig:tsne}
\end{figure}

\subsection{Alignment Visualization}

To qualitatively assess the effect of OSA alignment, we compare t-SNE visualizations
before and after alignment on 100 randomly sampled ML-1M interactions,
colored by rating level (Figure~\ref{fig:tsne}). Before alignment, representations exhibit no clear organization with respect to
preference. After OSA alignment, interaction representations are organized according to the ordinal structure of the rating anchors. Interactions with
adjacent ratings are positioned closer, while distant ratings
are less likely to overlap, reflecting the relative ordering encoded in
the anchors.
By aligning collaborative representations into the same semantic space as
the anchors, OSA enables the LLM to interpret and utilize both CF signals and fine-grained
preference differences during recommendation.

%% file: texs/4.Conclusion.tex
\section{Conclusion}

In this paper, we address a limitation of existing CF--LLM hybrids that overlook
fine-grained user preferences.
OSA models preferences at the interaction level and aligns representations
to ordinal semantic anchors derived from explicit ratings,
preserving the ordinal structure of preference intensity.
Since these anchors are grounded in the LLM’s semantic space,
the model can naturally exploit its inherent understanding of graded preference signals
during recommendation.
Empirical results show that
this interaction-level, strength-aware preference modeling not only improves overall performance but
also better distinguishes subtle preference differences.
Future work may extend this framework by incorporating richer information to capture user preferences. One promising direction is to categorize implicit feedback into text-based semantic anchors, allowing the model to represent broader interaction patterns beyond explicit ratings.